\documentclass{aa}

\usepackage{graphics}

\begin{document}

\thesaurus{13.18.5,08.03.5,05.01.1,08.09.2}

\title{VLBI observations of two single dMe stars:\\ spatial resolution and astrometry}

\author{M. R. Pestalozzi \inst{1} \and A. O.  Benz \inst{2} \and J. E. Conway \inst{1} \and M. G\"udel \inst{3}}

\institute{Onsala Space Observatory, S-439 92 Onsala, Sweden \and Institute of Astronomy, ETH, CH-8092 Z\"urich, Switzerland \and Paul-Scherrer-Insitute, CH-5232 Villigen, Switzerland}

\date{Received 4 July 1999 / Accepted 24 September 1999}

\titlerunning{VLBI Observations of Two Single dMe Stars}
\authorrunning{Pestalozzi et al.}

\offprints{M.R. Pestalozzi, \\michele@oso.chalmers.se}

\maketitle

\begin{abstract}

We report on 3.6 cm VLA and VLBA observations of \object{YZ CMi} and \object{AD Leo}, two nearby dMe stars. We resolve YZ CMi and can fit a circular symmetrical gaussian component of FWHP of 0.98 $\pm$0.2 mas, corresponding to an extent of the corona above the photosphere of $1.77 \times 10^{10} \pm8.8 \times 10^{9}$ cm or 0.7 $\pm0.3$ $R_\star$ ($R_\star$ refers to the photospheric radius). We obtain an estimate of the brightness temperature of $7.3 \times 10^{7}$ K, which is consistent with that expected from gyrosynchrotron emission. For AD Leo the emitting region is unresolved. We therefore set a conservative upper limit to its diameter of 1.8 times the photosphere diameter, which leads to an extent of the corona above the photosphere of $<2.8\times 10^{10}$cm or $<0.8$ $R_\star$. We compare the radio emitting dMe stars with measured sizes with the Sun and conclude that these active stars have much more extended coronal radio emission than the Sun. The VLBA position of YZ CMi has been 
found to differ by 32 mas from the positions calculated from the Hipparcos 
catalogue. The discrepancy is caused by large errors in the listed proper motion. An improved value is given.


\end{abstract}

\section{Introduction}
\label{sec:uno}

Hot coronae are the enigmatic link between cool, convective stars and their 
environment. While most of this hot plasma ($>10^{6}$ K) is contained by magnetic fields, some may escape due to overpressure as a hot stellar wind along open field lines. The extent of intense coronal emissions in X-rays or radio waves thus yields a lower limit to the size of closed magnetic loops, i.e. the size of the stellar magnetosphere. 

Radio emission from {\it solar} active regions at 3 cm wavelength originates 
from altitudes above the photosphere between 5 to $10\times 10^8$cm, or 0.006 -
0.015 $R_\odot$, (e.g. \cite{Aschw} 1995). It is generally attributed to 
thermal gyroresonance emission. The size of the radio image of the Sun increases at longer wavelengths. In soft X-rays and EUV line emissions, coronal loops 
reach $3\times 10^{10}$cm (0.5 $R_\odot$) (e.g. \cite{Sturr}). Magnetic loops in excess of $1.4\times 10^{11}$cm (2 $R_\odot$) have been 
observed in metric U bursts (\cite{Labrum}). Coronograph observations 
in white light, however, do not regularly see loops of that size, suggesting 
that they are shortlived. In general, large magnetic loops are confined to low 
latitudes and, consequently, the radio image of the Sun is more extended in the equatorial direction.

Very Long  Baseline Interferometry (VLBI) has made it possible to study the size and shape of {\it main-sequence stellar coronae}. At 3.6 cm, \cite{B4} (1998) have resolved the dMe star UV Cet B at an higher and variable flux 
level into two components separated by $4.4\times 10^{10}$cm (4.4 $R_\star$). 
One of the components was spatially resolved with the size of about the stellar photosphere. The orientation of the two sources was found to lie along the probable axis of rotation, strongly suggestive of coronal enhancements extending at least $2.1\times 10^{10}$cm (2.1 $R_\star$) above the poles. Thus, UV Cet B differs considerably from the Sun in size and shape of the radio corona.

More VLBI observations of dMe stars have been reported at 18 cm wavelength. For YZ CMi, \cite{B1} (1991) found an upper limit of $8.7\times 10^{10}$cm, 
suggesting an extent of a circular corona above the photosphere of less than $1.8\times 10^{10}$cm (0.74 $R_\star$). Two size measurements of AD Leo by \cite{B2} (1995) suggest a coronal extent of less than $2\times 10^{10}$cm 
(1.0 $R_\star$) and $5.4\times 10^{10}$cm (2.6 $R_\star$) above the photosphere, assuming a circular, concentric shape. A small total size of less than 
$4.9\times 10^{10}$cm for EQ Peg (\cite{B2} 1995) refers to an  
observation during a totally polarized flare, and an extremely large size of 
$2.2\times 10^{11}$cm was reported for the dMe close binary YY Gem (\cite{Alef1}).

The radio coronae of active, rapidly rotating dMe stars have been modeled to 
determine the emission process. It is generally agreed (\cite{Manuel}; \cite{White}) that weakly polarized radio emission is produced by 
the gyrosynchrotron mechanism of a population of mildly relativistic electrons. This mechanism, however, cannot account for polarizations exceeding about 50\% which may originate from a coherent process. Circular polarizations of up to 80\% have been reported for YZ CMi at 20 cm and 18 cm (\cite{Lang}; \cite{B1} 1991). 

\begin{figure*}[!]
\resizebox{\hsize}{!}{\includegraphics{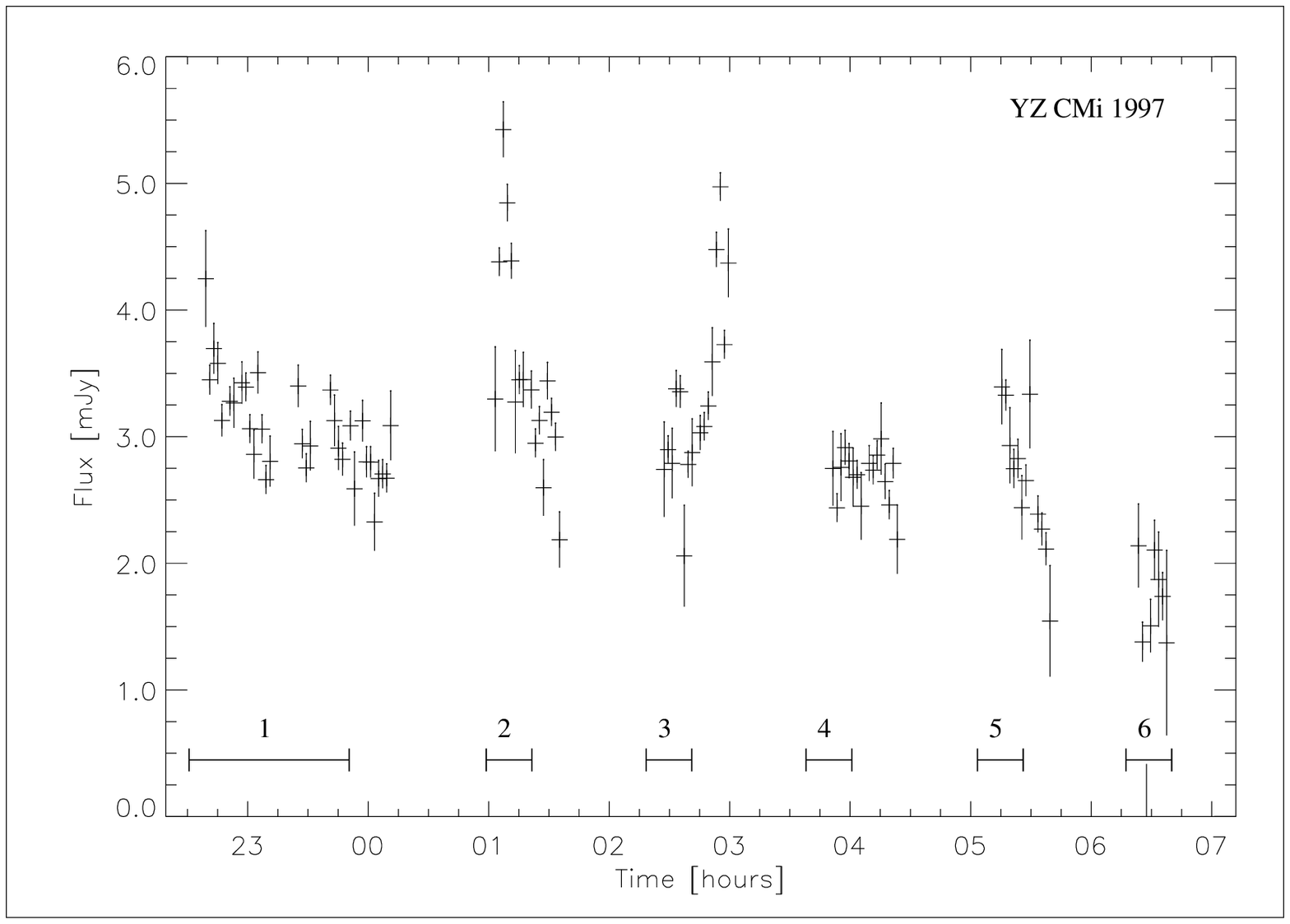}\includegraphics{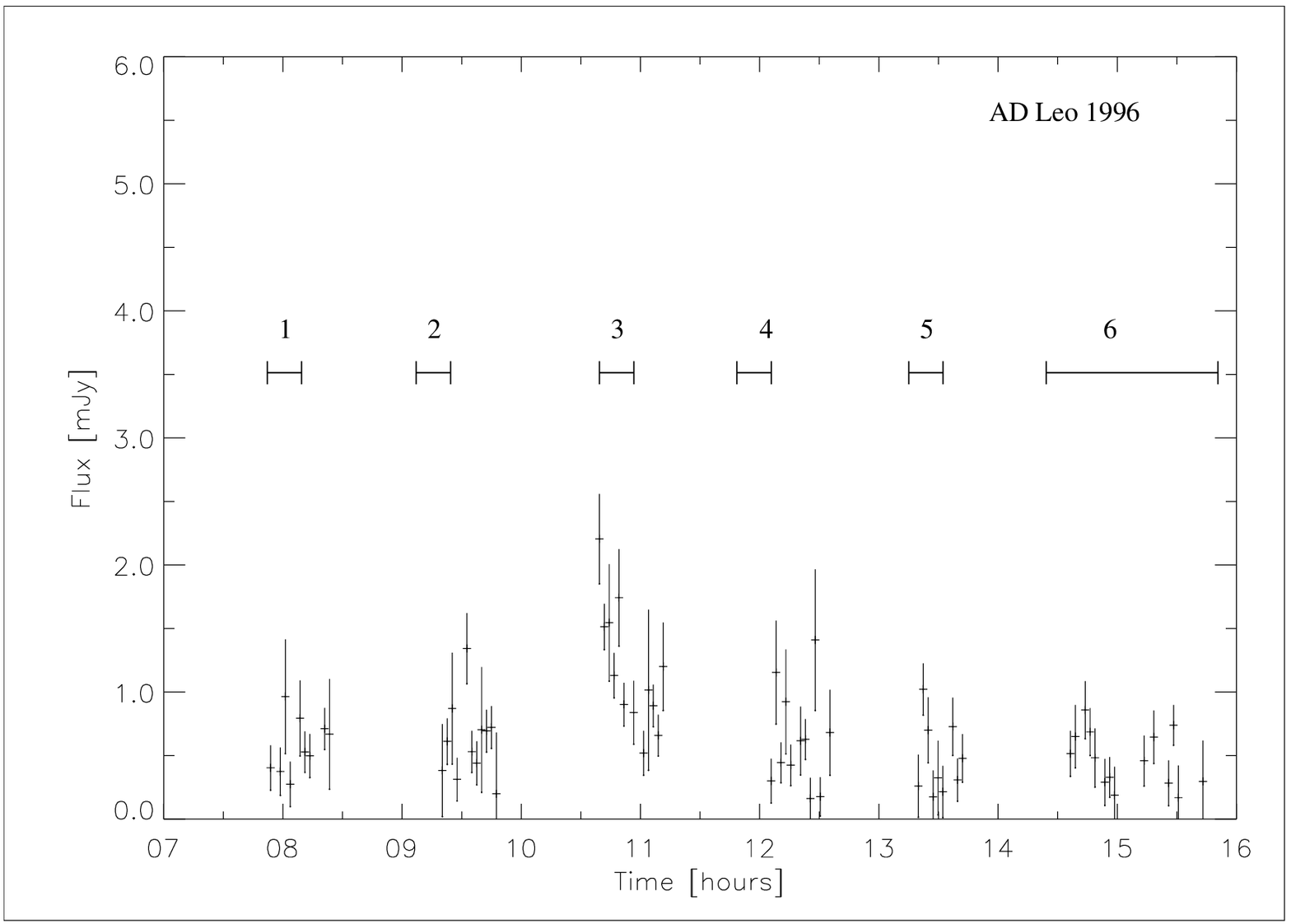} }
\caption{Total flux density (stokes I) against time during the YZ CMi and AD Leo observations, extracted from the VLA data. The relatively large average flux of YZ CMi is noticeable as are several flares. The horizontal numbered bars in the plot show the periods ({\it scans}) during which the VBLI data were collected. Each scan is composed of subscans on the target star and the calibrator alternately.}
\label{fig:light97}
\end{figure*}

Here we report on the results of VLBA experiments of YZ CMi and AD Leo at 3.6 
cm. These are  well known, nearby, young radio stars close to zero-age main 
sequence (ZAMS). Some of their general properties are listed in Table \ref{tab:sum}.

\section{Observations and data reduction}

The single dMe stars YZ CMi and AD Leo were observed on 1997 April 18/19 and 
1996 December 12, respectively, with the VLBA and the phased-up VLA as a joint 
VLBI system yielding an angular resolution of better than one milliarcsecond 
(mas). 

The VLA was also available in its normal interferometer 
mode and was used for total flux measurements at very small baseline length which allowed us to monitor the changes in total flux density and polarisation of both stars through the observations. 3C286 was used for the flux calibration of both stars.

The VLBI observations used a bandwidth of 8 MHz in both left and right circular polarization 
at 8.41 GHz (3.6 cm) and two bit sampling, giving a data rate of 128 Mbit/s. In order to reliably image such weak sources ($<3.0$ mJy) the phase referencing technique was used (\cite{Beasley}) in which we switched between the target and a bright calibrator (quasar) in cycles of two to three minutes. For our targets YZ CMi and AD Leo these calibrators were 0736+017 and 1022+194 respectively (see Table \ref{tab:cali}), at 2.1 and 1.4 degrees separation from the target. 

The amplitude calibration 
was performed with AIPS, followed by editing in DIFMAP (in order to flag more 
precisely single, bad visibilities of the VLBA data) and by a continued analysis in AIPS (including all the mapping). We first made hybrid maps of our calibrators using closure phase methods in order to determine their structure. We then determined the atmospheric contributions to the phase of the calibrator data. With these solutions it was possible to phase-correct the stellar data. We then made preliminary wide maps in order to find the stars and phase-rotated the data in order to bring the target stars to the image phase center. Finally we deconvolved the stellar images using CLEAN.

As a check of the reliability of the phase calibration, our observations also included a second bright calibrator for each star (0743-006 for YZ CMi, and 1013+208 for AD Leo, see Table \ref{tab:cali}). The maps of these secondary calibrators, made using the phase solutions found towards the primary calibrators showed a dynamic range of about 20:1. Since the separation between the two calibrators is approximately the same as that between the star and each calibrator, we can conclude that our stellar images have also a 20:1 dynamic range and hence are in practice noise limited rather than dynamic range limited. We also produced selfcalibration maps and datasets of the secondary calibrators. The plots of the fringe amplitude against {\it u,v}-distance show the same fall off than the ones produced from the phase referenced datasets. This confirms the value of the dynamic range of the images of the targets.

The source centroid positions of the phase referenced maps of the secondary calibrators were within 0.5 and 0.2 mas of the correlated positions (a-priori positions) for 0743-006 and 1013+208 respectively (see Table \ref{tab:cali}). This is consistent with the claimed accuracy of those correlated positions (0.5 mas, \cite{Johnston} 1995). This test gives us confidence in the astrometric accuracy of our stellar positions (see Sect. 3.4).

\section{Results}

\subsection{Total flux monitoring}

YZ CMi was found at a surprisingly high, slowly decreasing flux density (average  2.9 
mJy) during the whole observation (see Fig. \ref{fig:light97}, left plot). As previously noted (see Sect. \ref{sec:uno}), the emission of YZ CMi 
is often considerably polarized. Throughout this observation the polarization 
was predominantly left circular   (about 60 \% circular polarisation during the quiescent emission, and 90 \% during the flares). Two strong flares appear in the lightcurve obtained with the VLA. Their flux values reach up to 6.0 mJy. The detection of YZ CMi with both instruments was clear, as the rms noise in the VLA image was 
0.063 mJy/beam, and in our high sensitivity VLBI maps made using only VLBA-VLA baselines the noise was 0.13 mJy/beam. Given these map noise values the peak intensity of the star was 47 $\sigma$ in the VLA image, and 16 $\sigma$ in the VLBA image. 

The mean flux density value for AD Leo was at a more typical level of 0.7 mJy (see Fig. \ref{fig:light97}, right plot). Similar fluxes at 18 cm have previously been reported (e.g. \cite{Jack}; \cite{B2} 1995). A weak flare appears in the VLA lightcurve (2.1 mJy). The star was detected by both instruments despite its low flux. The rms noise for the VLA map was 0.035 mJy/beam, while for the map made with only the VLA-VLBA baselines it was 0.18 mJy/beam. Peak map values were 15.6 $\sigma$ for the VLA and 5 $\sigma$ for the VLBA maps respectively. Tracing the lightcurve of AD Leo from the interferometric VLA data was a delicate task because of two strong sources in the field of view.

\subsection{Size and shape of YZ CMi}

\begin{figure*}
\resizebox{\hsize}{!}{\includegraphics{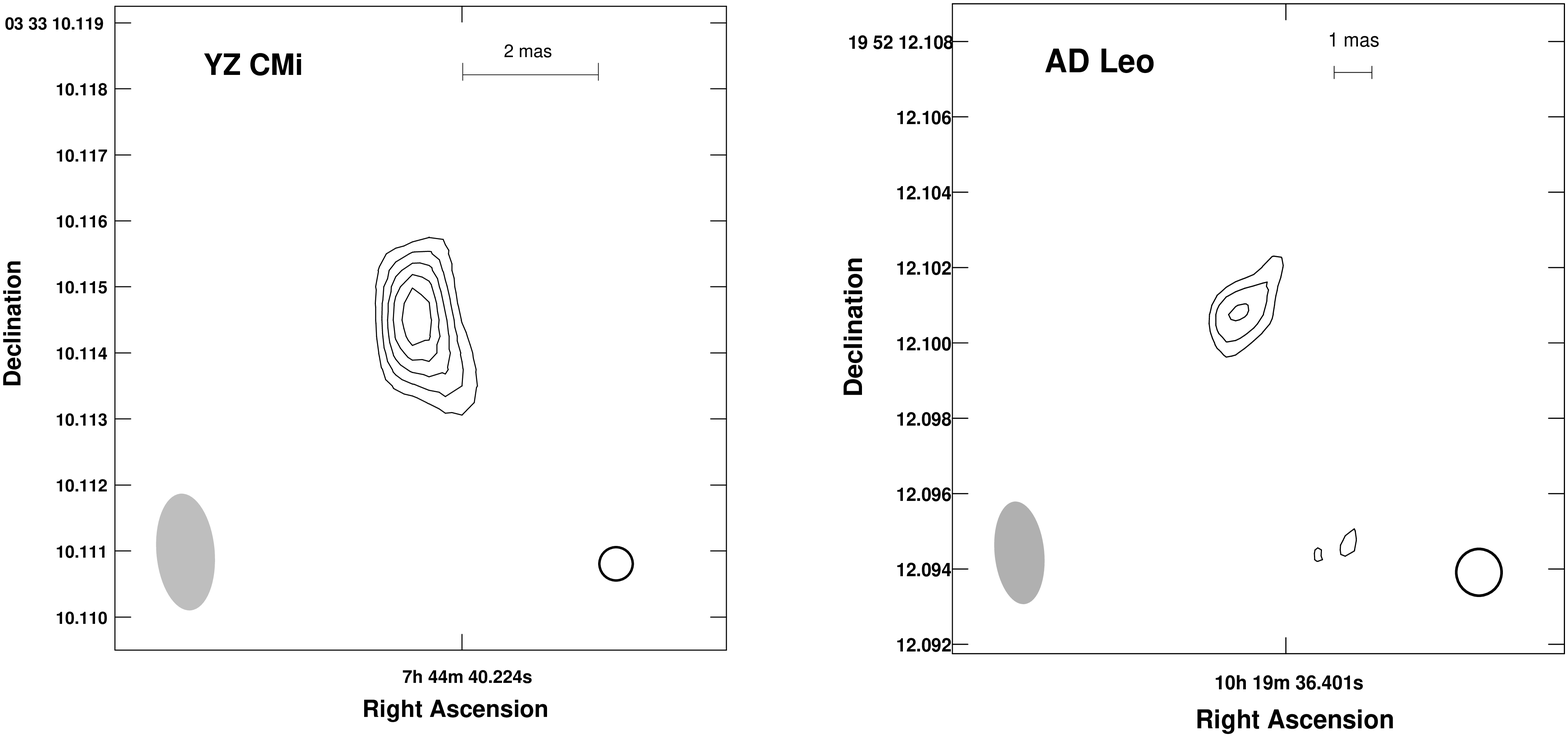}}
\caption{A contour plot of YZ CMi observed on April 18/19, 1997 (left) and AD Leo observed on December 12, 1996 (right). The maps have rms noise $\sigma$=0.13 mJy/beam and $\sigma$=0.18 mJy/beam respectively. The first 
contour is at 3 $\sigma$, and then at steps of one $\sigma$. At the bottom 
left of each panel the clean beam is drawn, and the circle in the lower right corner represents the expected optical size of the star, indicated in Table \ref{tab:sum}.}
\label{fig:cntr}
\end{figure*}

\begin{figure*}[!ht]
\resizebox{\hsize}{!}{\includegraphics{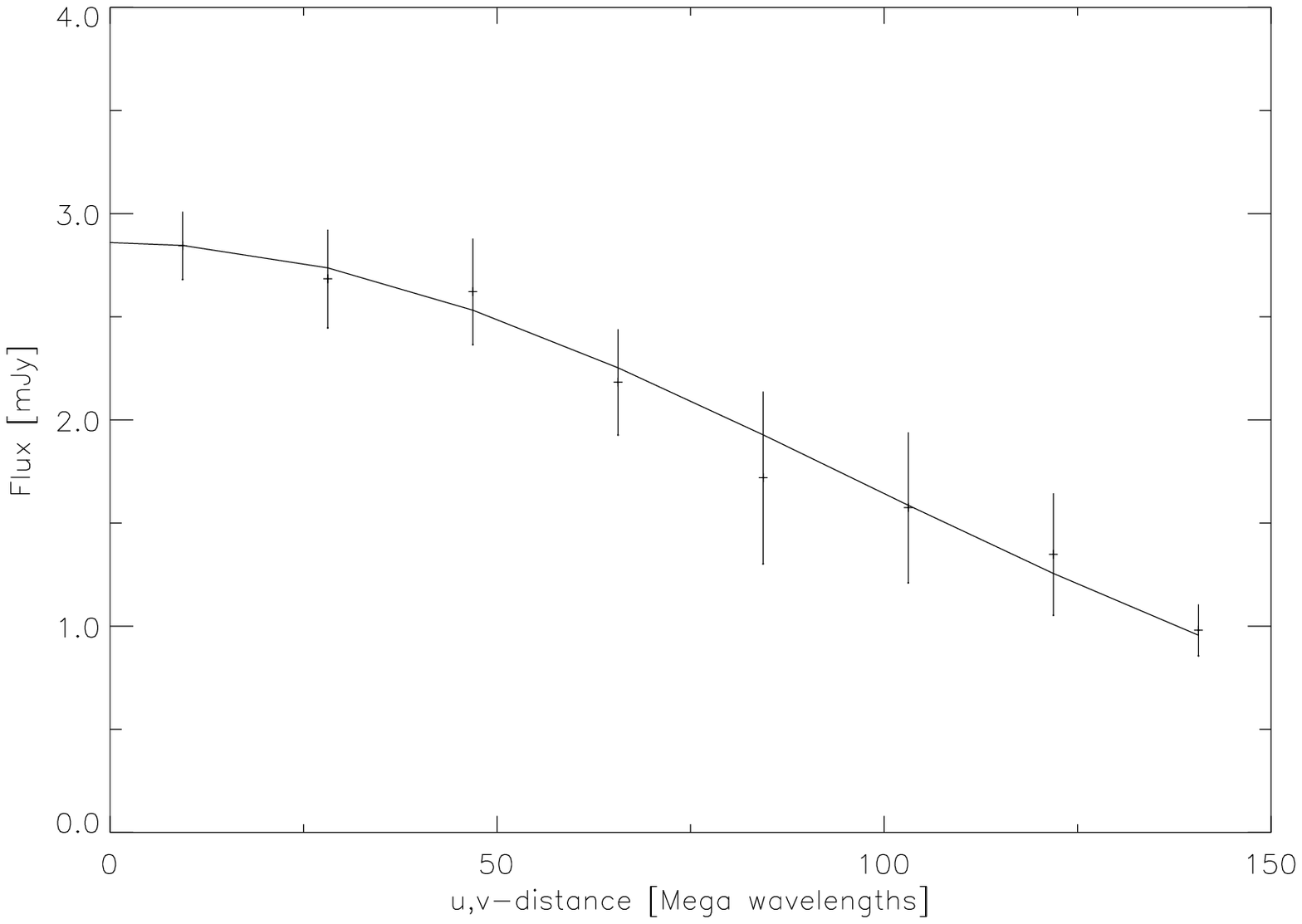}\includegraphics{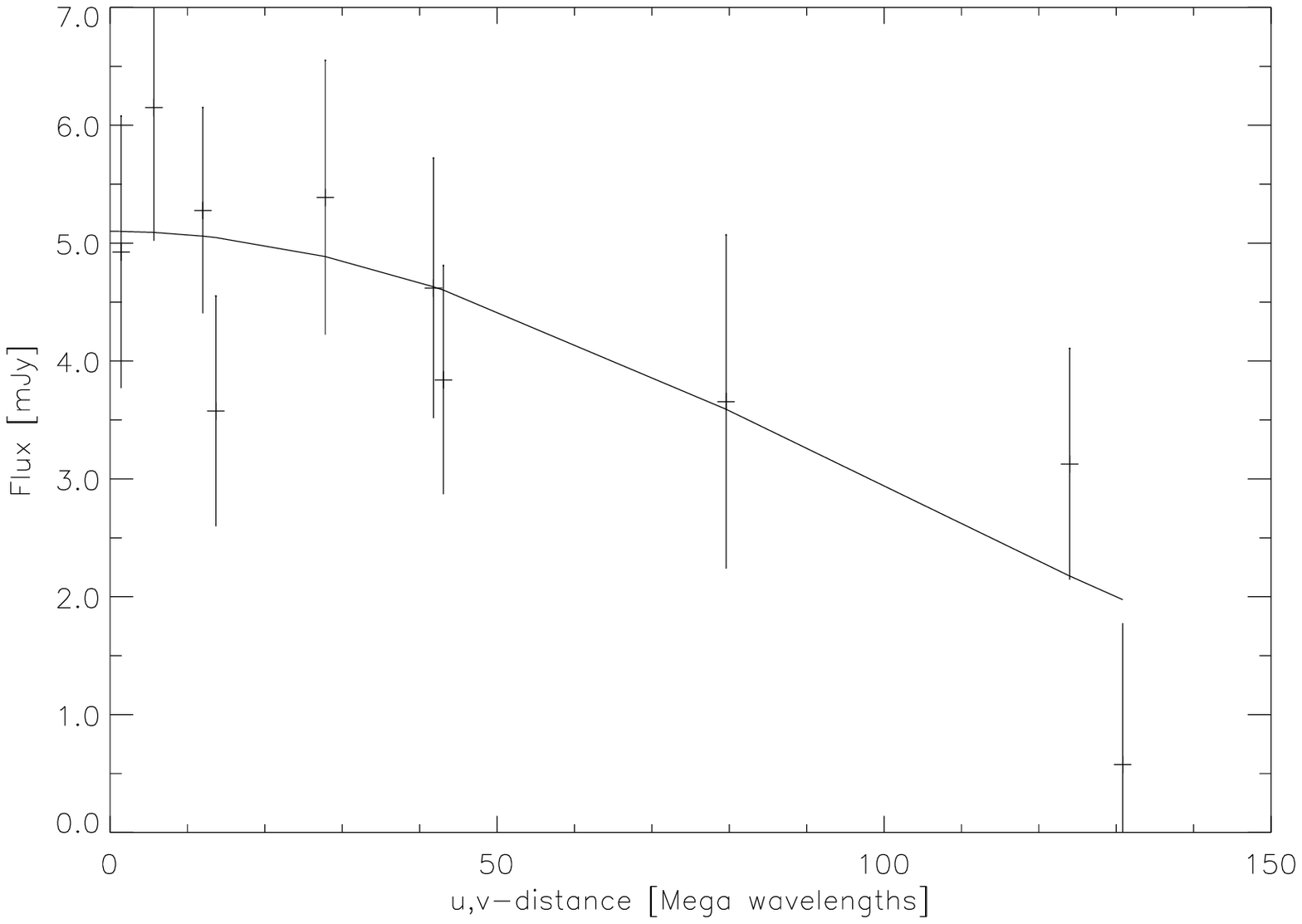}}
\caption{Fringe amplitude versus {\it u,v}-distance of YZ CMi on the sensitive VLA-VLBA baselines. The solid line is the result of fitting a circular gaussian (see also Table \ref{tab:sum}). Note the change in vertical scale of the right plot compared to the left one. The right pannel shows the amplitude versus $u,v$-distance during the strong flare in scan 2. The solid line in this plot represents the same gaussian model used in the left plot except for the total flux density value which has been increased to 5.1 mJy.}
\label{fig:uvplots}
\end{figure*}

The phase referenced VLBA map of YZ CMi is shown in Fig. \ref{fig:cntr}. Evidence for spatial resolution was found in this map. Further evidence for extended emission can be most clearly seen by examing fringe amplitude versus baseline length obtained using the VLA-VLBA baselines (see Fig. \ref{fig:uvplots}, left plot). We obtained this plot by first coherently averaging the data on each baseline-scan (75 minutes) in order to increase the signal to noise ratio significantly above unity. We then binned the amplitude over baseline length finding the mean amplitude in each bin by incoherent averaging. The error bars on this average were determined from the internal scatter of the data.

The fall off noticeable in Fig. \ref{fig:uvplots} left plot clearly indicates a resolved source. What is more, the phase values on VLBI baselines to the phased VLA over the whole observation show no significant variation from zero. There is therefore no evidence for anything other than a single centro-symmetric component.
We searched also for other evidence for non-symmetrical structure looking at closure phases, but the SNR of these were too low. 

Given the close to quadratic fall off of the fringe amplitude with \emph{u,v}-distance shown in Fig. \ref{fig:uvplots} left plot it is impossible to distinguish between gaussian, sphere, disk or ring like models (\cite{Pearson}). We therefore fitted one-component gaussian models to the YZ CMi data. The numerical values for the fits are summarized in Table \ref{tab:sum}. The dimensions of sphere, disk and ring which would show similar fits are expected to be respectively 1.8, 1.6 and 1.1 times the gaussian FWHP values (Pearson 1995). Two ways of fitting were followed: fitting in AIPS using the task UVFIT and our own model fitting to the data. The first fitted an elliptical gaussian and obtained for the whole data set a major axis of FWHP of 1.4 $\pm$0.3 mas and a minor axis of 0.5 $\pm$0.25 mas (1$\sigma$ errors). There is therefore no strong evidence for ellipticity, and our subsequent fitting of the data outside of AIPS assumed only a circular gaussian. With such a model it was possible to fit most of the data within 1 $\sigma$. The best fitting FWHP size of the corona was found to be 0.98 $\pm$0.2 mas, which corresponds to 1.7 $\pm$0.3 stellar diameters.

\begin{table*}[!ht]
\begin{center}
\begin{tabular}{lll}
\hline \hline 
 & YZ CMi 1997 & AD Leo 1996 \\
\hline \\
Spectral type & M 4.5Ve & M 4.5Ve \\
Distance & 5.93 $\pm$0.09 pc & 4.90 $\pm$0.07 pc \\
Optical diameter & 0.58 mas & 0.95 mas \\
\hline
(UVFIT all obs) & 1.4x0.5 ($\pm$0.3x0.25) mas &  \\
 & PA 12.8$^o$ ($\pm$5.1) &  \\
(gaussfit all obs) & 0.98 ($\pm$0.2) mas &  \\
(UVFIT scan 2) & 0.890 ($\pm$0.22) mas & \\
(gaussfit scan 2) & 1.02 ($\pm$0.2) mas & \\
pm and plx & $\Delta\alpha$: -0.32 mas & $\Delta\alpha$: -0.90 mas\\
 during observations & $\Delta\delta$: -0.16 mas & $\Delta\delta$: +0.29 mas \\
\hline
VLBA Position & $\alpha$: 7$^{h}$ 44$^{m}$ 40\fs 22399 $\pm$0.00007 & $\alpha$: 10$^{h}$ 19$^{m}$ 36\fs 40112 $\pm$0.00007 \\
 (J2000) & $\delta$: 3\degr 33\arcmin 10\farcs 1140 $\pm$0.0005 & $\delta$: 19\degr 52\arcmin 12\farcs 1008 $\pm$0.0005 \\
Mean flux density & & \\
during observation & 2.97 mJy & 0.54 mJy \\
Brightness temperature & 7.29$\cdot 10^{7}$ K & $>4.93\cdot 10^{7}$ K \\
\hline
\end{tabular}
\caption{Summary of the results from the observations and fits for YZ CMi and AD Leo. The sizes of the radio emission are labeled with the name of the procedure used 
to obtain them (UVFIT in AIPS, gaussfit outside AIPS). The distances are taken from the 
Hipparcos and Gliese catalogues, respectively, and the optical diameters are 
from \cite{Pettersen} (1980). For the AIPS fit of scan 2 we constrained the program to fit a circular gaussian because of the small number of visibilities. We also did a fit for AD Leo but it did not converge (see text).}
\label{tab:sum}
\end{center}
\end{table*}

Since the second VLBI scan corresponds precisely to one of the two strong flares (see Fig. \ref{fig:light97}, left plot), it was interesting to study it more closely. Fig. \ref{fig:uvplots} (right plot) shows the amplitude versus {\it u,v}-distance. Of this scan we selected the subscan on the target with the highest flux density value and coherently averaged the data over it (3 minutes), obtaining one point per VLA-VLBA baseline. The solid line corresponds to the same gaussian model fitted to the whole data set (see Fig. \ref{fig:uvplots}, left plot) except for the total flux density which was increased to 5.1 mJy. We find that this scaled model fits the data within the errors and therefore there is no evidence for a change in source size during the flare. Independent gaussian fits to the data are also consistent with this conclusion (see Table \ref{tab:sum}).

We should add that the contribution of the 
proper motion and of the changing parallax of the star during the ten hours of 
observation is -0.32 mas and -0.16 mas in $\alpha$ and $\delta$, respectively. 
These values are small enough that they do not give a significant contribution on the spatial extent.

\begin{table*}
\begin{center}
\begin{tabular}{cccc}
\hline \hline
 Star & Name of calibrator & $\alpha$(J2000) & $\delta$(J2000) \\
\hline
YZ CMi & 0763+017 & 07$^{h}$ 39$^{m}$ 18.0339$^{s}$ & 01$^{0}$ 37$^{'}$ 04.619$^{''}$ \\
 & 0743-006 & 07$^{h}$ 45$^{m}$ 54.0823$^{s}$ & -00$^{0}$ 44$^{'}$ 17.538$^{''}$ \\
\hline
AD Leo & 1022+194 & 10$^{h}$ 24$^{m}$ 44.8096$^{s}$ & 19$^{0}$ 12$^{'}$ 20.416$^{''}$ \\
 & 1013+208 & 10$^{h}$ 16$^{m}$ 44.3251$^{s}$ & 20$^{0}$ 37$^{'}$ 47.336$^{''}$ \\
\hline
\end{tabular}
\caption{Adopted positions for the correlation of the primary and secondary calibrators of each target star. The primary calibrators were used as reference sources in the phase referencing, the secondary calibrators as check sources (see Sect. 2). All positions come from \cite{Johnston} (1995).}
\label{tab:cali}
\end{center}
\end{table*}

\subsection{AD Leo}

An image of AD Leo is shown in Fig. \ref{fig:cntr}. It appears slightly elongated. This image is probably affected by the star's high proper motion (Table \ref{tab:sum}), since the extension 
is exactly along the expected direction: the star appears blurred because of its motion during the synthesis observation.

As shown in the flux curve of AD Leo (Fig. \ref{fig:light97}, right plot), the total flux varied by a factor of 3 during the observations. Adding this fact to the high proper motion of the star, it is not surprising that attempts to fit a single gaussian to the $u,v$-data did not give consistent results. The weakness and variability of the star make the estimation of errors on the size difficult. However, from the image (see Fig. \ref{fig:cntr}), we can note that the apparent FWHP perpendicular to the motion corresponds to the beam FWHP in this direction. The intrinsic FWHP size of the emitting region is therefore likely to be less than half the beam FWHP or about 1 mas, which equals the estimated optical diameter of the star (see Table \ref{tab:sum}). This might indicate a very compact corona or an emitting spot on the surface of the star. A very conservative upper limit on the size of the corona in the former case would be to assume the emitting region to be an optically thin sphere instead of a gaussian (\cite{Pearson}) in which case the diameter is less than 1.8 times the photosphere diameter, and it therefore has an extent above the photosphere of less than 0.8 $R_\star$.

\begin{table*}[!ht]
\begin{center}
\begin{tabular}{lllll}
\hline \hline 
star &radio luminosity & extent of corona above&\hskip-3mm photosphere & 
reference\\
  &[erg/s\ Hz] & cm & $R_\star$ &\\
\hline \\
Sun & $5.7\times 10^{10}$ & $7.4\times 10^8$ & 0.011 $R_\star$ & \cite{Aschw} 1995 \\
AD Leo & $3.04\times 10^{13}$ & $< 2.8\times 10^{10}$ & $< 0.8 R_\star$ &this 
work \\
UV Cet B & $8.8\times 10^{13}$ & $2.1\times 10^{10}$ & 2.1 $R_\star$ & \cite{B4} 1998 \\
YZ CMi & $1.4\times 10^{14}$ & $1.77\times 10^{10}$ & 0.7 $R_\star$ &this work \\
\hline
\end{tabular}
\caption{Luminosity and coronal size  of Sun and dMe stars at 3.6 cm 
wavelength.}
\label{tab:size}
\end{center}
\end{table*}

\subsection{Astrometry}

The relatively large signal-to-noise ratio for YZ CMi and the phase referencing to a calibrator with good ($<$0.5 mas) positional accuracy in the radio frame (\cite{Johnston} 1995) allowed us to determine a 
precise position for this star. This position was compared with the position given by the Hipparcos catalogue (ESA 1997). Correcting for proper motion and parallax, we found a discrepancy of 20.9 mas in $\alpha$ and 30.4 mas in $\delta$, thus a total deviation of 36.9 mas. The proper motion of the star is given in the Hipparcos catalogue as -344.9 $\pm$2.6  mas/yr in $\alpha$, and -450.8 $\pm$1.75 mas/yr in $\delta$. Considering the time interval between the two measurements of 6 years, the difference is 2 $\sigma$ and thus within the accuracy of the proper motion error bars. Combining the VLBA and Hipparcos positions (courtesy of F. Arenou), an improved proper motion of -348.6 $\pm$0.6 mas/yr in $\alpha$, and -446.6 $\pm$0.3 mas/yr in $\delta$ can be derived.

The position of AD Leo obtained with the VLBA was compared with those 
available in the Gliese and Tycho catalogues. The latter showed a deviation with the VLBA position of 176.3 mas and 100.0 mas in $\alpha$ and $\delta$, 
respectively. They are within one standard deviation of the Tycho catalogue 
accuracy.

\section{Discussion and conclusions}

VLBA observations have spatially resolved YZ CMi and the data could be fitted with a circular gaussian of a FWHP of 0.98 $\pm$0.2 mas. The radio corona 
extent is $1.77\times 10^{10} \pm8.8\times 10^{9}$ cm above the photosphere (the phospheric radius is assumed to be $2.6\times 10^{10}$cm, \cite{Pettersen} 1980). For AD Leo, which is closer 
and has a larger photosphere, but was observed at a much weaker flux level, the corona was not resolved and we set a robust upper limit of $2.8 \times 10^{10}$ cm above the photosphere (see Sect. 3.3).

\subsection{The brightness temperature}

For YZ CMi (see Table \ref{tab:sum}) we obtain a mean brightness temperature $T_{b}=7.3 \times 10^{7}$ K, while for AD Leo we set an upper limit of $T_{b}=4.93 \times 10^{7}$ K. These mean $T_{b}$ values are smaller than previously reported for YZ CMi by \cite{B1} (1991) and AD Leo at 18 cm (\cite{Jack}).
The lower values at 3.6 cm are still consistent with a non-thermal spectrum from gyrosynchrotron but do not formally exclude thermal processes. However the significant circular polarisation found during the observations strongly argues for a gyrosychrotron emission mechanism.

The derived extent of the coronae above the photosphere of the dMe stars is compared to the Sun in Table \ref{tab:size}. The solar value refers to stereoscopic measurements of the thermal gyroresonance emission of active regions. The average value at 10-14 GHz reported by \cite{Aschw} (1995) has been used.

The experience from the solar radio emission makes it clear that the observed 
radio size is only a lower limit of the size of the stellar corona. Nevertheless, these results indicate that the observed active dMe stars have much larger active coronae than the Sun. It might be that such dMe stars posess systems of closed loops reaching heights in excess of a stellar diameter. One way to realize such extended coronae is by large distances between footpoints as possibly seen in the case of UV Cet B (\cite{B4} 1998). This indicates either that active regions are very large, or that active loops preferentially connect different active regions.

\begin{acknowledgements}
We thank F. Arenou for the combined evaluation of VLBA and Hipparcos astrometry data. The Very Large Baseline Array and the Very Large Array are operated by 
Associated Universities, Inc. under contract with the US National Science 
Foundation. The work at ETH Zurich is financially supported by the Swiss 
National Science Foundation (grant No.  20-046656.96 ).
\end{acknowledgements}

\bibliographystyle{astron}

\end{document}